\documentclass[%
 reprint,
 amsmath,amssymb,
 aps,
]{revtex4-2}
\usepackage{graphicx}
\usepackage{dcolumn}
\usepackage{bm}
\usepackage{color} 
\usepackage{cancel}
\usepackage{array}
\usepackage{tabularx}
\usepackage{booktabs}
\usepackage{graphicx}
\usepackage{dcolumn}
\usepackage{braket}

\usepackage{booktabs}

\makeatletter
\newcommand{\tpmod}[1]{{\@displayfalse\pmod{#1}}}
\makeatother

\begin{document}

\title{Symmetry-enforced metal-insulator transition and topological adiabatic 
charge pump in sliding bilayers of threefold symmetric materials}

\author{Sergio Bravo}
\email{sergio.bravoc@usm.cl}
\address{Departamento de F\' isica, Universidad T\'ecnica
Federico Santa Mar\' ia, Av. Espana 1680, Casilla Postal
110V,Valpara\' iso, Chile}

\author{P. A. Orellana}
\address{Departamento de F\' isica, Universidad T\'ecnica
Federico Santa Mar\' ia, Av. Espana 1680, Casilla Postal
110V,Valpara\' iso, Chile}

\author{L. Rosales}
\address{Departamento de F\' isica, Universidad T\'ecnica
Federico Santa Mar\' ia, Av. Espana 1680, Casilla Postal
110V,Valpara\' iso, Chile}

\date{\today}

\begin{abstract}
Sliding bilayers are systems that exploit the possibility of relatively
translating two monolayers along a specific direction in real space, such that
different stackings could be implemented in the process. This 
simple approach allows for manipulating the electronic properties of 
layered materials similarly as in twisted multilayers. In this work, the
sliding of bilayers, composed of one type of monolayer with spatial symmetry
described by space group P$\bar{3}1m$ is studied. Using a minimal
tight-binding model along with symmetry analysis, we propose two effects that
arise in a specific sliding direction.
First, the sliding-induced control of the band gap magnitude, which produces a 
metal-insulator transition, is demonstrated. In addition, the potential to
achieve a topological adiabatic charge pump for cyclic sliding is discussed.
For each effect, we also present material implementations using 
first-principles calculations. Bilayer GaS is selected for the metal-insulator
transition and bilayer transition metal dichalcogenide ZrS$_2$ is found
to display the topological pump effect. Both realizations show good agreement
with the predictions of the model.
\end{abstract}

\maketitle

\section{Introduction \protect\\} 

The advent of two-dimensional materials composed of stacked atomic layers, 
has opened an immense spectrum of possibilities to realize new physical
effects and also novel applications based on the reduced spatial 
dimensionality and the associated quantum phenomena that this implies.
The configuration of these systems allows for precise control
of the layer-by-layer composition, such as in Van der Waals
heterostructures \cite{Geim2013,Cast2022} and also permits an effective
inclusion of external perturbations such as gating, light, and
magnetic fields among others \cite{VdW_adma}.

A complementary form of structural manipulation relates to the relative spatial
orientation of the constituent layers. In general terms, we have two 
possible scenarios: relative rotation and relative translation between layers. 
The cases of the first group, commonly known
as twisted multilayer systems, comprise the materials where some
monolayers are rotated with respect to a reference state such that a
fixed point in space is maintained in the process. 
On the other hand, when a subgroup of layers is rigidly translated
relative to the rest of the constituents, a sliding process is
established. In this work, we concentrate on sliding configurations for 
the particular case of bilayer materials.

Bilayer systems have recently been the focus of intense research that have
uncovered tantalizing physical properties, such as superconductivity and
correlated phases in twisted bilayer graphene \cite{Cao2018,Cao2018_2} and
charge ordered phases along with Wigner crystal states in
hexagonal transition-metal chalcogenides \cite{Mak2022}. 

Concerning sliding bilayer systems, interfacial sliding have been
realized experimentally in several TMDCs, including MoS$_2$, MoSe$_2$,
WS$_2$, WSe$_2$ and WTe$_2$ \cite{Wang2022,acschemrev_slide}.
Theoretical studies where the implications of sliding are 
explored include nanoscale friction in TMDCs 
\cite{nanofrict_TMD_2017}, quantum phase 
transitions in MnBi$_2$Te$_4$ \cite{QPT_MnBi2Te4_2022},
circular dichroism in bilayer graphene
\cite{slipandtwist_2019}
and topological effects in sliding bilayer graphene 
\cite{slide_graph_2011}, twisted bilayer graphene
\cite{topchar_TBG_2020,Moire_edge_TBG_2021} and other 
Moiré structures 
\cite{topcharge_slidemoire_2020,top_slide_moire_hetero_2020}.

In this work, we explore bilayer systems composed of one monolayer type. 
The spatial symmetry of the monolayer is described by the
space group P$\bar{3}1m$ (\#164). In turn, the symmetry of the sliding
bilayer system can preserve the space group of the parent monolayer for
particular stacking configurations or (in general) can decrease the
symmetry to a subgroup of space group \#164. 

In what follows, we describe the physical consequences that sliding
can produce in these bilayers. We start in section \ref{II} with a generic
description of the geometry of the systems in terms of space groups and 
we put forward a minimal tight-binding model to describe the sliding
process along a particular spatial direction in terms of a single scalar
parameter. This model allows us to introduce two possible effects  
that arise from the sliding event. Namely, a metal-insulator transition 
that is enforced by specific symmetry and filling constraints 
owing to the transformations that affect the
symmetry representations in momentum space.   
Also, for an adiabatic cyclic sliding setting, the possibility 
of achieving a topological charge pump along the periodic directions 
of the lattice. 

Guided by the model results, we present in section \ref{III} a material
realization for each effect employing first-principles calculations,
which provides numerical support for the feasibility of the proposed
phenomena. We conclude in section \ref{IV} by commenting on the
prospects for implementation and the possible extensions that can be
put forward. Additional information complementing the results 
presented in the main text has been left as 
electronic supplementary material (SI).

\begin{figure*}
\includegraphics[width=1.9\columnwidth,clip]{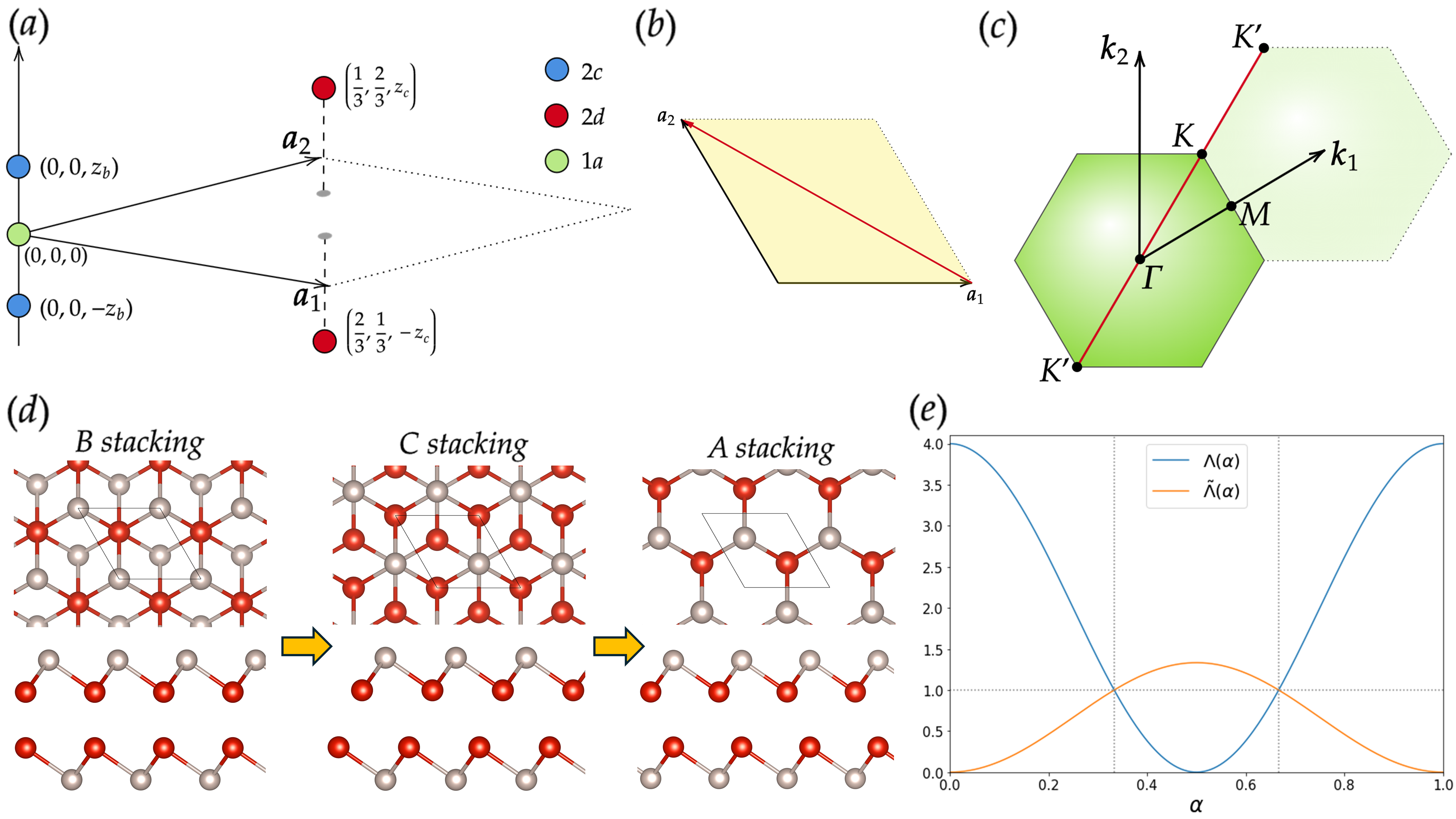}
\caption{(a) Unit cell and Wyckoff positions for the systems used in
this work. 
(b) Real space displacement used for sliding. (c) Two-dimensional 
Brillouin zone. (d) lattice structures for the high-symmetry
stackings that arise along the sliding process. (e) Variation of the
tight-binding interlayer hopping parameters for the internal
atoms as a function of the $\alpha$ parameter.}
\label{FIG-1}
\end{figure*}

\section{Geometric description and tight-binding model}\label{II}
\subsection{geometric configurations}

As we mentioned above, the bilayer systems studied are conformed by a monolayer
having a lattice structure described by space group (SG) P$\bar{3}1m$ (\#164).
We will use this SG \#164 as our reference for symmetry analysis since the
bilayer stackings of interest can preserve this SG or reduce it to a subgroup
of SG \#164. This SG is generated by a threefold rotation around the axis
perpendicular to the bilayer plane, a twofold rotation along an axis within
the bilayer plane, and spatial inversion (in three dimensions). 
To describe all the configurations we will use a unit cell defined
in three-dimensional space by the vectors $\boldsymbol{a}_1 = a(1,0,0)$, 
$\boldsymbol{a}_2 = a(-1/2,\sqrt{3}/2,0)$, where $a$ is the in-plane lattice
constant, as shown in FIG. \ref{FIG-1}a. We include the out-of-plane
coordinate $z$ to account for the nonzero thickness of the structures since
atomic positions have nonzero $z$. The same unit cell is maintained for all
the stackings that compose the sliding process. 

We describe the sliding as a rigid relative translation between the two
monolayers, characterized by a single vector in the plane formed
by $\boldsymbol{a}_1$ and $\boldsymbol{a}_2$ vectors. In this work, we focus
on particular sliding paths that contain stackings that have their symmetry
described by SG \#164. In general, three stackings fulfill this
requirement, the features of which will be detailed below. Guided by this 
high symmetry stackings (HSS), we find that six possible sliding
paths contain all three HSS along the displacement. The vectors that
define the direction of these paths are given by 

\begin{equation*}
\begin{split}
    & \pm (\boldsymbol{a}_1 + \boldsymbol{a}_2), \\
    & \pm (\boldsymbol{a}_1 - \boldsymbol{a}_2),  \\
    & \pm (\boldsymbol{a}_1 - 2\boldsymbol{a}_2).
\end{split}
\end{equation*}

In what follows, we will focus on one of these possible 
translations, namely $\boldsymbol{a}_1 - \boldsymbol{a}_2$,
since the other alternatives do not add new features to the effects
discussed in this work. The selected sliding vector is depicted 
in FIG. \ref{FIG-1}b. 
The three formerly introduced HSS will be defined in terms of a minimal
lattice model where only four occupied atomic positions are used. 
In practical terms, we keep the bottom layer of the system fixed and 
change the top layer coordinates to simulate the sliding process in
general. The bottom layer has coordinates (in terms of the lattice
vectors $\boldsymbol{a}_1$, $\boldsymbol{a}_2$)

\begin{equation}
    \mathbf{r}_{1,b} = (1/3,2/3,-z_{i}) , \mathbf{r}_{2,b} = (2/3,1/3,-z_{e}) 
\end{equation}
where $-z_{i}$ and $-z_{e}$ denote the fractional $z$ coordinate of the 
atoms in bottom monolayer, such that $z_{i}$ corresponds the top atom
in the monolayer, thus being an internal atom of the bilayer and
$z_{e}$ corresponds to the bottom atom in the monolayer, becoming
an external atom in the bilayer.

The first stacking, which we denote as stacking A, will have the top layer
coordinates
\begin{equation}
    \mathbf{r}_{1,t} = (1/3,2/3,z_{e}) , \mathbf{r}_{2,t} = (2/3,1/3,z_{i}). 
\end{equation}
The next HSS, stacking B, possesses coordinates
\begin{equation}
   \mathbf{r}_{1,t} = (0,0,z_{e}) , \mathbf{r}_{2,t} = (1/3,2/3,z_{i}). 
\end{equation}
Finally, stacking C is defined as
\begin{equation}
   \mathbf{r}_{1,t} = (2/3,1/3,z_{e}) , \mathbf{r}_{2,t} = (0,0,z_{i}). 
\end{equation}

We can now see how these three HSS can be obtained along the sliding path.
If we parameterize the relative translation of the top layer by 
$\alpha(\boldsymbol{a}_1 - \boldsymbol{a}_2)$, where $\alpha$ is a scalar 
that can take values from 0 to 1, then general coordinates for the top layer
can be defined by 

\begin{equation}
\begin{split}
    &\mathbf{r}_{1,t}(\alpha) = (-\alpha,\alpha,z_{e}) , \\
    &\mathbf{r}_{2,t}(\alpha) = (1/3-\alpha,2/3+\alpha,z_{i}). 
\end{split}
\label{coor_alpha}
\end{equation}
Therefore, we get the sequence of stackings B,C,A,B for $\alpha=\{0,1/3,2/3,1\}$,
respectively. The lattice structures for this sequence of stackings are pictured 
in FIG \ref{FIG-1}d. When $\alpha$ takes other values, we can use the same
expression in eq. (\ref{coor_alpha}) to describe the generic sliding stage. 
It is worth mentioning that all the non-HSS stackings along the paths described
above belong to the SG \# 12. This SG can be considered as the 
group of the sliding path, a concept that is known as the distortion symmetry
group \cite{VanLeeuwen2015}. Coincidentally, inversion symmetry is conserved
along the whole path, while the threefold symmetry is only preserved in 
the HSS. This symmetry breaking for general stackings is responsible for
the fact that degeneracies at special points in the momentum space can
be controlled in the process.

Although all the HSS belong to the same SG \#164, there exist differences in the
symmetry classification for each stacking within this SG. The key ingredients 
are the Wyckoff positions (WP) that are occupied for each configuration. 
A bilayer system in this SG, in the HSS with the above atomic positions, 
will have two possible WP, $2c$ and $2d$. A schematic representation of the
location of this WP in the unit cell is presented in FIG. \ref{FIG-1}a. 
It has to be considered that the coordinates of the WP as presented in
this figure will not directly correspond with the coordinates for each HSS. 
In order to have a perfect match for all HSS with the coordinates of the WP
we should have to redefine a new unit cell for each case. However, in our case,
we conserve the same unit cell because this enables a unified treatment of the
sliding by using one parameter. Thereby, we only consider the underlying
feature of each WP, that is the general relation of the coordinates. For the $2c$ WP
the relation is that they have mirror symmetry about the bilayer plane and that
their in-plane coordinates are the same. For the case of $2d$ WP, the atomic 
positions must be connected by one of the vectors $\pm (-1/3,1/3,2\tilde{z})$,
where $\tilde{z}={z_{i},z_{e}}$. With this identification, we classify the WP for 
the three HSS. A summary of this procedure is presented in TABLE \ref{TAB1}.

The utility of this WP identification is that it will provide a more in-depth
understanding of the electronic effects derived from sliding, and will also guide
the tight-binding model, which we proceed to outline next.

\begin{table}
\begin{tabular}{@{}ccc@{}}
\toprule
Stacking & 2c & 2d \\ \midrule
A        & $(\frac{1}{3},\frac{2}{3},z_{e}),(\frac{2}{3},\frac{1}{3},-z_{e})$ &    -  \\
\vspace{1mm}
         & $(\frac{1}{3},\frac{2}{3},z_{i}),(\frac{2}{3},\frac{1}{3},-z_{i})$ &       \\
\vspace{1mm}
B        & $(\frac{1}{3},\frac{2}{3},z_{i}),(\frac{1}{3},\frac{2}{3},-z_{i})$ &
$(0,0,z_{e}),(\frac{2}{3},\frac{1}{3},-z_{e})$ \\
C        & $(\frac{1}{3},\frac{2}{3},z_{e}),(\frac{1}{3},\frac{2}{3},-z_{e})$ &
$(0,0,z_{i}),(\frac{2}{3},\frac{1}{3},-z_{i})$ \\ \bottomrule
\end{tabular}
\caption{Identification of the Wyckoff positions for each high-symmetry stacking.}
\label{TAB1}
\end{table}

\subsection{Minimal tight-binding model}

The starting point for developing the minimal model is considering
that we have four atomic positions at hand. Thus, we will decorate each position
with a single spinless orbital. From symmetry considerations, each atomic position
belongs to a unique WP, as explained in the previous paragraph. 
A so-called site-symmetry group (SSG) can be associated to each WP position, which comprises
the spatial symmetries that leave the WP points invariant \cite{TQC2017}. 
This SSG is important since it allows classifying the atomic orbitals
according to the irreducible representations (irreps) of 
the corresponding group \cite{TQC_AR_2021}. Finally, when passing to the momentum
space description of the system, a set of orbitals (transforming as a SSG irrep),
will induce a set of irreps at every point in the Brillouin zone (BZ) \cite{TQC2017}.
This set of irreps coming from the atomic limits in real space is known as a
elementary band representation (EBR) \cite{TQC_PRB_2018}.
These EBRs give the symmetry content in reciprocal space and yield information on
how the wavefunctions must transform along the whole BZ. 

For the bilayers in this work, we have two WP, and each WP contains two
single-valued irreps, $A{1}$ and $E$, following the notation of the Bilbao 
Crystallographic Center (BCS) \cite{Bilbao_2017}. Each irrep will induce a
specific set of irreps in reciprocal space. The information of this map
from WP to momentum space symmetry is presented in TABLE S1 in the SI. 

More concretely, as we use four atomic orbitals, we will obtain a 
four-band model with a 4$\times$4 Hamiltonian. We will use as a basis set
orbitals that transform as the $A_{1}$ irrep for each WP ($s$ , $p_{z}$ or $d_{z^2}$), 
which we will denote and order as $\ket{\psi_{t,e}},\ket{\psi_{t,i}},\ket{\psi_{b,e}},
\ket{\psi_{b,i}}$. The first subscript indicates if the state belongs to the bottom 
($b$) or top ($t$) layer and the second if it is associated with an internal ($i$)
or external ($e$) site of the bilayer. The next step is to define the degree of
hopping interactions that will be included in the model. To keep the parameters at
a minimum, we include the first
and second nearest-neighbor (NN) intralayer interactions and two direct interlayer 
interactions, one among the internal atoms and the other for 
external atoms of the bilayer. In this manner, the Hamiltonian matrix of the system
will be expressed as

\begin{multline}
H =  \\
\begin{bmatrix}
E_e+H_{intra}^{(2,e)} & H_{intra}^{(1)} & H_{inter}^{(i)} & 0 \\
H_{intra}^{(1)*} & E_i+H_{intra}^{(2,i)} & 0 & H_{inter}^{(e)} \\
H_{inter}^{(i)*} & 0 & E_e+H_{intra}^{(2,e)} & H_{intra}^{(1)} \\
0 & H_{inter}^{(e)*} & H_{intra}^{(1)*} & E_i+H_{intra}^{(2,i)} 
\end{bmatrix}
,
\end{multline}
where $E_i$ and $E_e$ are the onsite energies for the internal and external
bilayer sites, respectively. On the other hand, each hopping matrix element can
be defined in the following way. The first NN intralayer interaction 
$H_{intra}^{(1)}$ takes the form

\begin{equation}
    H_{intra}^{(1)} = t_1(e^{i\mathbf{k}\cdot\mathbf{\delta_1}}+
    e^{i\mathbf{k}\cdot\mathbf{\delta_2}}+e^{i\mathbf{k}\cdot\mathbf{\delta_3}}),
\end{equation}
where $\mathbf{k}$ is the reciprocal space vector, $t_1$ is the hopping amplitude
and the $\delta_{j}$ represent the first intralayer NN vectors.
The second NN intralayer interactions for internal and external atoms,
$H_{intra}^{(2,\mu)}$ ($\mu=\{e,i\}$), is given by

\begin{equation}
    H_{intra}^{(2,\mu)} = t_{2,\mu}(e^{i\mathbf{k}\cdot\mathbf{\gamma_1}}+
    e^{i\mathbf{k}\cdot\mathbf{\gamma_2}}+e^{i\mathbf{k}\cdot\mathbf{\gamma_3}}),
\end{equation}
such that $t_{2,\mu}$ is the hopping amplitude and the second
intralayer NN vectors are denoted as $\gamma_{j}$. The interlayer hopping between
internal atoms of the bilayer, $H_{inter}^{(i)}$, can be expressed as

\begin{equation}
    H_{inter}^{(i)} = \Lambda_{i}(\alpha) e^{i\mathbf{k}\cdot\mathbf{\xi_1(\alpha)}}+
    \tilde{\Lambda}_{i}(\alpha)(e^{i\mathbf{k}\cdot\mathbf{\xi_2(\alpha)}}
    +e^{i\mathbf{k}\cdot\mathbf{\xi_3(\alpha)}}).
\end{equation}
In this last expression the $\xi_{j}$ represent the internal interlayer NN
vectors while $\Lambda_{i}(\alpha)$ and $\tilde{\Lambda}_{i}(\alpha)$ denote
the interlayer hopping amplitude among these atoms as a function of the sliding
parameter $\alpha$. Lastly, the interlayer interaction among the external atoms of
the bilayer can be written as

\begin{equation}
    H_{inter}^{(e)} = \Lambda_{e}(\alpha)e^{i\mathbf{k}\cdot\mathbf{\zeta_1(\alpha)}}+
    \tilde{\Lambda}_{e}(\alpha)(e^{i\mathbf{k}\cdot\mathbf{\zeta_2(\alpha)}}+
    e^{i\mathbf{k}\cdot\mathbf{\zeta_3(\alpha)}}),
\end{equation}
where $\Lambda_{e}(\alpha)$ and $\tilde{\Lambda}_{e}(\alpha)$ are the alpha-dependent
hopping amplitudes for this interaction and the $\zeta_{j}$ denote the corresponding
external interlayer NN vectors.

The explicit form of all the NN vectors defined above, intralayer and interlayer, is 
included in the SI.

The $\Lambda_{\mu}(\alpha)$ ($\mu=i,e$) amplitudes vary their magnitude as the sliding is 
produced. It is thus necessary to establish an explicit functional form in order to 
obtain a parametric Hamiltonian. We model the amplitudes following symmetry rules that
are dictated from the HSS. In the first place, for the internal amplitudes, at the B stacking
($\alpha=0$), $\Lambda_{i}(\alpha)$ must be maximal and $\tilde{\Lambda}_{i}(\alpha)$ must
be zero. In addition, for the C ($\alpha=1/3$) and A ($\alpha=2/3$) stackings, 
$\Lambda_{i}(\alpha)=\tilde{\Lambda}_{i}(\alpha)$.
In the case of the external amplitudes, the conditions are that, at C stacking, 
$\Lambda_{e}(\alpha)$ must be maximal and $\tilde{\Lambda}_{e}(\alpha)$ must be zero.
While for both, A and B stackings, $\Lambda_{e}(\alpha)=\tilde{\Lambda}_{e}(\alpha)$. 
Regarding the remaining (non-HSS) stackings, which interpolate between the above cases,
we only impose that they give a smooth variation of the amplitudes with respect to 
$\alpha$. 
We propose the following modulated functional forms that meet all the above criteria.

\begin{equation}
    \begin{split}
    \Lambda_{\mu}(\alpha)  & = 4 \lambda_{\mu} \cos^2(\pi\alpha-\phi_{\mu}), \\
    \tilde{\Lambda}_{\mu}(\alpha) & = \frac{4}{3} \lambda_{\mu} 
    \sin^2(\pi\alpha-\phi_{\mu}),
    \end{split}
\end{equation}
where $\phi_\mu=\{0,\pi/3\}$ for $\mu=\{e,i\}$, respectively. The
$\lambda_{\mu}$ scalar parameters are defined as; the internal hopping amplitude of
the C and A stackings for $\lambda_{i}$ and the external hopping amplitude of the A
and B stackings for $\lambda_{e}$. An example of the general form of the hoppings as
a function of $\alpha$ is presented in FIG. \ref{FIG-1}e for the internal interactions. 
The external hoppings behave in a similar manner only with a change of 
phase with respect to the internal interactions.
We now put into use the model developed and show how it gives raise to two effects
which depend on the sliding process and also on the symmetry character of the
bands.

\subsection{Metal-insulator transition} \label{II.C}

In the first place, we study how the sliding process can control the gap between 
energy bands when we connect the B stacking to the C stacking by sliding, that
is when alpha takes values from 0 to $\pi/3$.
The reason for selecting these two stackings is that they comprise the same    
type of WP, one $2c$ and one $2d$, which will make the effect more transparent. 
More in detail, we analyze the symmetry content that the orbitals induce in
momentum space. As we have four A$_1$ orbitals, inspection of TABLE S1
indicates that, if the A$_1$ orbital locates at a $2c$ WP then the irreps
induced at the high-symmetry points of the BZ correspond to the set
($\Gamma_{1}^{+}(1)$ $\oplus$ $\Gamma_{2}^{-}(1)$, 
$K_{1}(1)$ $\oplus$ $K_{2}(1)$, $M_1^{+}(1)$ $\oplus$ $M_2^{-}(1)$)
(we follow the notation of the BCS \cite{Bilbao_2017}, see FIG
\ref{FIG-1}c for a representation of the momentum space high symmetry points).
The numbers in parenthesis denote the dimensionality of the irrep. 
Conversely, if the orbital is located at the $2d$ WP, the set of irreps at 
the high-symmetry points will be the
($\Gamma_{1}^{+}(1)$ $\oplus$ $\Gamma_{2}^{-}(1)$, $K_3(2)$,
$M_1^{+}(1)$ $\oplus$ $M_2^{-}(1)$). It can be noted that the main
difference is at the $K$ point, where for the $2c$ WP, two 1-D irreps are
possible, while for the $2d$ WP, only one 2-D irrep is available. 
These two sets give all the irreps at our disposal to label the four bands
of the model. For illustrative purposes, we fix a particular order and
connectivity of the irreps to set a schematic representation for the band
structure. The resulting diagram will be 

\begin{equation*}
\begin{array}{ c c c }
\Gamma _{2}^{-},K_{2},M_{2}^{-} &  & \Gamma _{2}^{-},K_{2},M_{2}^{-}\\
\Gamma _{1}^{+} \oplus \Gamma _{2}^{-},K_{3},M_{1}^{+} \oplus M_{2}^{-}
& \rightarrow  
& \Gamma _{1}^{+},K_{1},M_{1}^{+}\\
\Gamma _{1}^{+},K_{1},M_{1}^{+} &  & \Gamma _{1}^{+} \oplus 
\Gamma _{2}^{-},K_{3},M_{1}^{+} \oplus M_{2}^{-}
\end{array}
.
\end{equation*}
Here, the B stacking is represented by the left group of irreps and the 
C stacking corresponds to the right group of irreps.
Therefore, if we, for example, set a filling of 1/4 such that only the
lowest-lying band is occupied at B stacking, it can be immediately noted
that in passing to the C stacking, a gap-closing process must take place.
This is because the lowest-lying band for this case corresponds to a
connected group of two bands, which for the current filling must realize
a gapless phase. In summary, if there exists an adequate band filling
along with the correct band ordering and
symmetry, it is possible to control the bandgap of the bilayer by means of sliding, 
producing a metal-insulator transition. We demonstrate the effect at the
tight-binding level with a band structure calculation for the B and C stackings 
for  a particular set of model parameters. The results are shown in 
FIG. \ref{FIG-2}a. Note that the path used in the band plot is defined as a
red line in FIG. \ref{FIG-1}c. Also, the Fermi level has been set to zero to
make evident the gap closing effect due to the change in stacking. In the next
section, we give a more detailed description of the sliding-induced metal-insulator
transition employing first-principles methods for a material example.

\subsection{Topological adiabatic charge pump}\label{II.D}

The second situation that we want to explore is related to the evolution of the 
polarization of the system when a complete sliding cycle is carried from 
$\alpha=0$ to $\alpha=1$. It is recalled that we cover all 
stackings in this process. 
To calculate the polarization of the system for every value of the sliding 
parameter, we use the modern theory of polarization in terms of the Berry phase 
as presented for example in \cite{polari_1993,vanderbilt_2018}. 
For the sliding bilayers we treat in this work, we have a two-dimensional
reciprocal space (see FIG. \ref{FIG-1}c) and in addition
the sliding parameter. Thus, a three-dimensional parameter space $(\alpha,k_x,k_y)$
is needed to analyze the polarization. The definition for component $j$
($j=\{1,2\}$) of the polarization, denoted as $P_j$, is given by \cite{vanderbilt_2018}

\begin{equation}
P_{j} =\frac{-e}{A_{cell}}\sum_{n}^{occ}\frac{\bar{\phi}_{n,j}}{\pi}
\ \boldsymbol{a}_{j},
\end{equation}
where $\bar{\phi}_{n,j}$ is the Berry phase for a band $n$ along direction $j$
averaged over the complementary direction in reciprocal space $l$. Explicitly

\begin{equation}
    \bar{\phi}_{n,j}=\frac{1}{2\pi}\int \phi_{n}^{k_{j}}(k_{l}) dk_{l},
\end{equation}
where $l=\{1,2\}$ and $\phi_{n}^{k_{j}}(k_{l})$ is the one-dimensional 
Berry phase along direction $j$ for a fixed value of $k_l$, which is
defined as \cite{RevModPhys_Berry}

\begin{equation}
\phi_{n}^{k_{j}}(k_{l}) =\ i\int \langle u_{n\boldsymbol{k}}|
\partial _{kj} u_{n\boldsymbol{k}}\rangle dk_{j}.
\end{equation}
In this last expression $u_{n\boldsymbol{k}}$ represents the unit cell 
periodic part of the eigenvector for the $n$-th band.  

The calculation of the polarization as presented above can be performed
separately for each value of $\alpha$ in the interval $[0,1]$. In order 
to study the evolution of the polarization along the sliding, we have to
take into account that the set of bands under study must be isolated in
energy from the rest of the bands in the spectrum for all values of 
$\alpha$. In our four-band model, the minimal set of bands that can 
meet this condition is a two-band subset, and for simplicity, we chose 
to set the filling to one-half in such a manner that only these 
low-lying bands contribute to the polarization. Therefore, we choose
a new set of tight-binding parameters. We have checked that 
the low-lying bands are gapped with respect to the upper bands 
throughout the sliding process. 

\begin{figure}
\includegraphics[width=\columnwidth,clip]{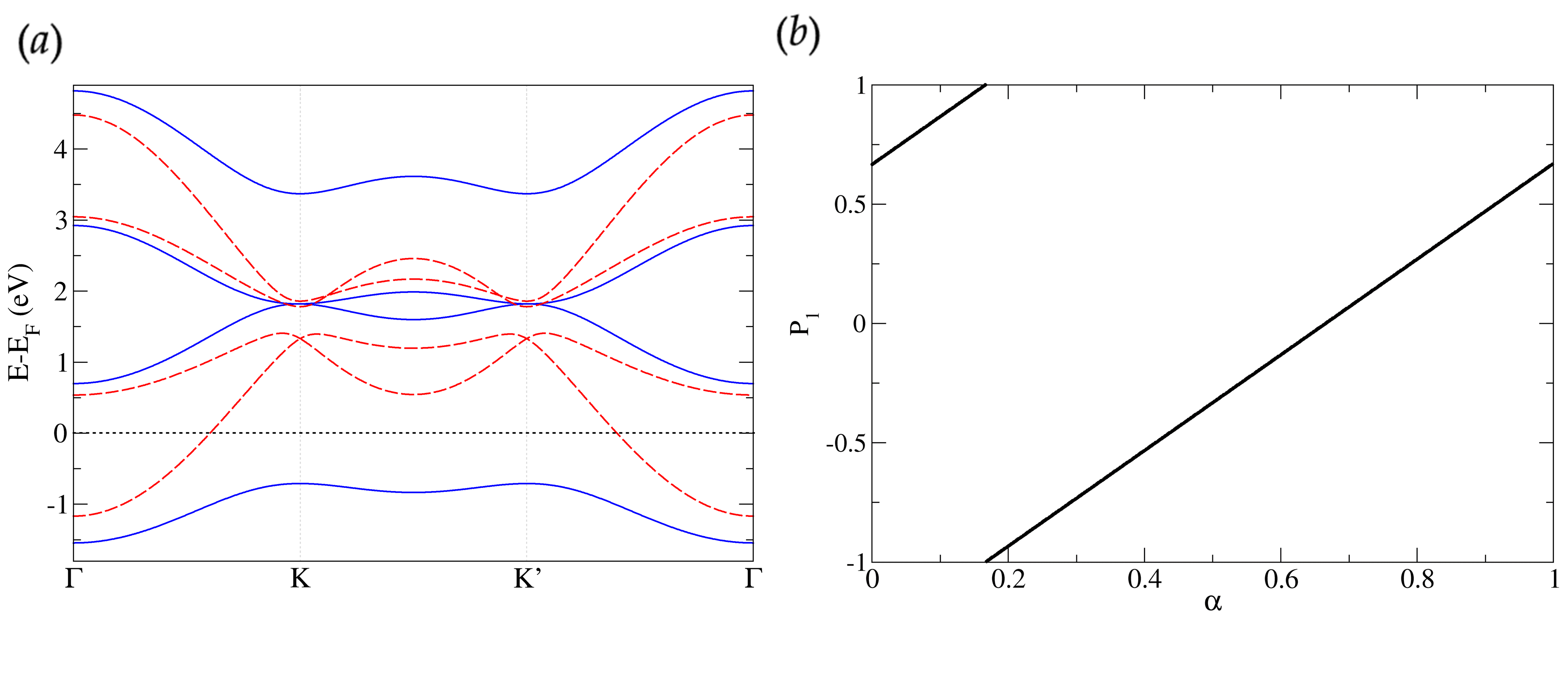}
\caption{(a) Band structures for the tight-binding model (used in section 
\ref{II.C}) with parameters
$E_{e}=1.85$, $E_{i}=1.40$, $t_1=0.63$, $t_{2,e}=0.01$, $t_{2,i}=0.023$,
$\lambda_{i}=0.51$, $\lambda_{e}=0.01$.
(b) Polarization, in units of the spinless quantum of polarization, 
as a function of $\alpha$ for the tight-binding model 
(used in section \ref{II.D}) using parameters 
$E_{e}=0.75$, $E_{i}=-0.75$, $t_1=1.0$, $t_{2,e}=0.09$, $t_{2,i}=-0.12$,
$\lambda_{i}=-0.15$, $\lambda_{e}=0.01$.}
\label{FIG-2}
\end{figure}

The calculation of the polarization can be numerically implemented
(see the Appendix for details) from the above formulas, and a complete
cycle of sliding can be studied such that $P_j(\alpha=0)=P_j(\alpha=1)$.
This implies that we are analyzing a closed surface in the 
$(\alpha,k_1,k_2)$-space, and as such, we can associate a Chern number to 
the evolution of the polarization \cite{RevModPhys_Berry,vanderbilt_2018}
along direction $j$, which we denote as $C(j)$.
The calculation of the polarization as a function of $\alpha$ for
$j=1$ is presented in FIG. \ref{FIG-2}b. It is evident that this two-band
polarization presents a nontrivial winding. As it is well known, the 
Chern number can be associated with the degree of the winding that
the polarization shows as a function of the external parameter 
\cite{topchar_TBG_2020}. 
Thus, for this minimal set of (two) bands, polarization winds once, 
and in consequence the Chern number along direction $j=1$ is $C(1)=1$.
This nonzero Chern number implies that the sliding along the 
special direction presented here entails a topological adiabatic 
process, which in principle, will allow us to observe a charge pump effect
along direction $\mathbf{a}_1$ in real space. An analogous calculation
of the polarization along direction $j=2$ yields a Chern number of 
$C(2)=-1$, which implies a topological charge pump along the direction 
$\mathbf{a}_2$. 
From the results of this model, we can infer that a pair of bands
that get connected for some value(s) of $\alpha$ will contribute 
a Chern number with magnitude $|C(j)|=1$. Then, in a system with
$N$ bands, we can expect the magnitude of the Chern number
to be $|C(j)|=N/2$.
Although spatial inversion is present in all the steps of the sliding,
it is still possible to define a charge pump effect since polarization 
is a lattice-valued quantity and can take nonzero values in 
inversion-symmetric if the set of values respect the underlying symmetry
\cite{vanderbilt_2018}. 
This implies that the variation of this lattice of values with respect to
the sliding parameter allows for the adiabatic charge pump effect.

Acquainted with the previous results at the tight-binding level, 
in the following, we are going to provide a realization of each
phenomenon in representative bilayer materials.

\section{Materials examples}\label{III}
\subsection{Metal-insulator transition in bilayer GaS}

As a sample system to implement the first effect, we will use a
two-layer material made of two monolayers of GaS. The lattice structure
has the same configuration as the model worked previously,
namely, the two Ga atoms comprise one WP, and the S atoms are located at the
other WP (see FIG. \ref{FIG-1}d). The Ga atoms are the internal atoms
and the S atoms form the external atom set. This material has been 
synthesized experimentally in layered form and present promising 
optoelectronic properties
\cite{Raman_GaS_2020,2D_CMDB_2021,PhysChemC_GaS_2021,
ACS_NLO_GaS_2022,ACS_app_2023_GaS}.
First-principles calculations were carried to simulate each step of the 
sliding. The details of the calculation are presented in the Appendix.
To provide evidence of the control of the gap, we follow the same 
approach as in the tight-binding model and compute the band 
structure as a function of $\alpha$. The results are depicted for
a representative array of alpha values in FIG. \ref{FIG-3}c, connecting
the stacking B to stacking C. The metal-insulator transition is clearly
identified and is achieved at the value of alpha $\alpha \approx 0.2$ and 
remains closed until stacking C is reached.

To explore the energetic landscape of the sliding process, 
the energy variation, $\Delta E$, of the ground state for a 
particular stacking with respect to the stacking with lower energy 
(stacking B in this case) is displayed in FIG. \ref{FIG-3}b, for a complete
cycle of $\alpha$. The cost of sliding from B to C is about $30$ meV. It can
be noted that the most costly part of the cycle is to slide from C to A, 
whose value reaches approximately $60$ meV. 
If the complete cycle is considered, the gap must be reopened on 
approaching stacking B from stacking A. This gives in principle
a cyclic metal-insulator transition controlled by the sliding. 
Band structure calculations for a selected set of $\alpha$ values 
which account for the range between C and A stackings is presented 
if FIG. S1 and for the range between A and B stackings in FIG. S2
in the SI.

As discussed in the context of the tight-binding model, the transition
robustness stems from the symmetry and ordering of the irreps at momentum 
space. For the GaS case, at the B stacking the top valence band is
nondegenerate and realizes the $\Gamma _{1}^{+},K_{1},M_{1}^{+}$ band
representation. The first and second conduction bands, for their part, 
realize the connected set 
$\Gamma _{1}^{+} \oplus \Gamma _{2}^{-},K_{3},M_{1}^{+} \oplus M_{2}^{-}$.
Then the low energy region of GaS could be well described by the model
we discussed above, when only the lowest band is filled. 
At B stacking, the top valence band can be considered as the bonding band
of the set coming from the $2c$ WP (Ga atoms) while the antibonding band 
$\Gamma _{2}^{-},K_{2},M_{2}^{-}$, will be higher in energy, locating above
the bands coming from the $2d$ WP. The filling restricts these $2d$-related 
bands to be antibonding conduction bands.
Thereby, as the WP get reversed at stacking C, the low-lying bands will
be enforced to form a connected two-band representation 
(having the 2-D $K_3$ irrep at $K$ point) which leads to a partially filled
bonding band, thus protecting the transition appearance.
Interestingly, the band representations induced from $2c$ WP in the
B and C stackings can be decomposed in EBRs pinned to $1a$ WP 
(see \ref{FIG-1}a), which is an unoccupied WP. This indicates that
these configurations realize a bilayer spinless obstructed atomic phase 
\cite{TQC2017,HOTCI_PRB_2019,OAI_PRB_2023}, 
which in particular is insulating at the B stacking.

\subsection{Topological adiabatic charge pump in bilayer ZrS2}

\begin{figure*}[ht]
\includegraphics[width=1.9\columnwidth,clip]{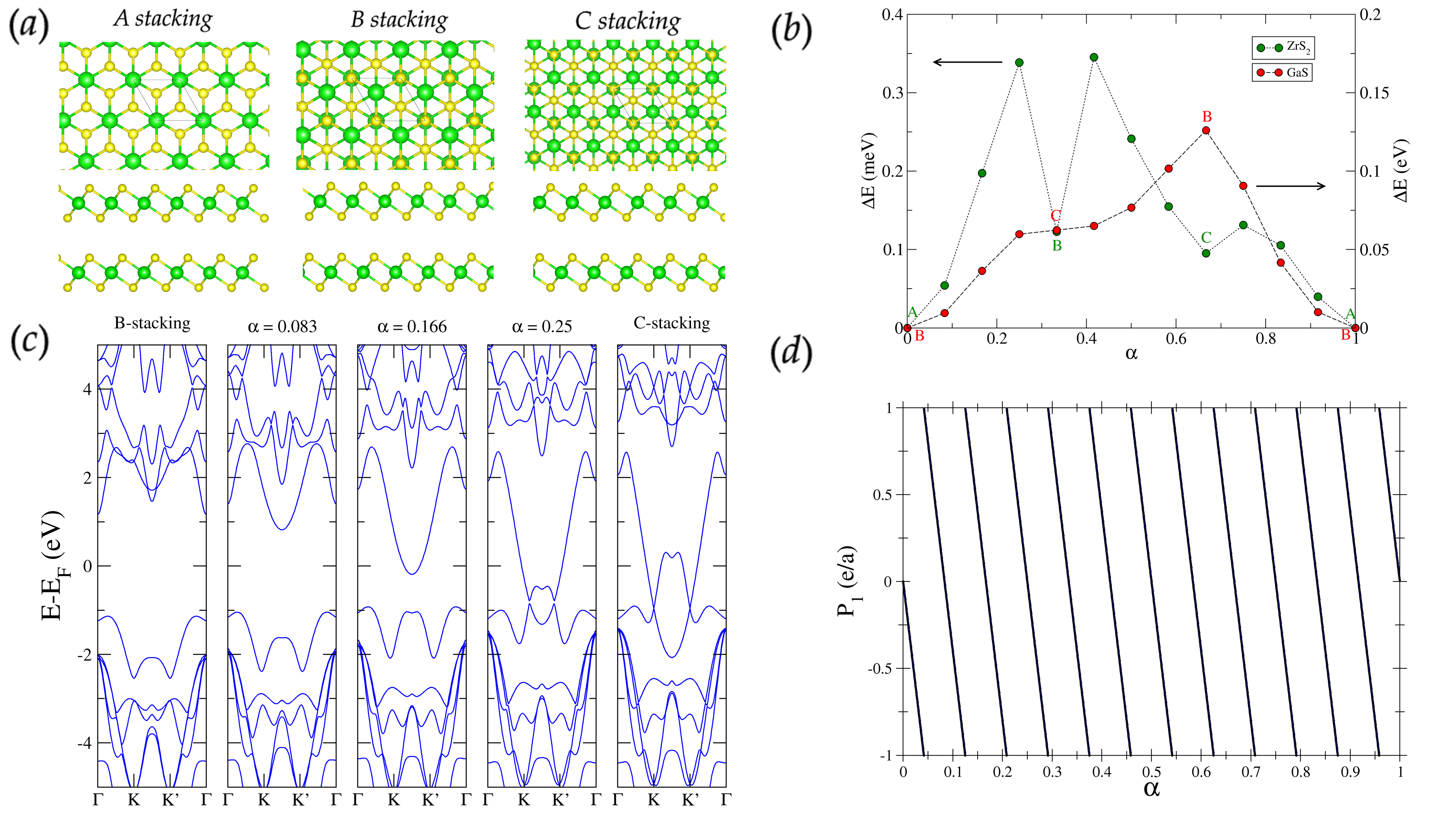}
\caption{(a) Lattice structures for the high-symmetry stacking in bilayer ZrS$_2$.  
(b) Energy difference $\delta$E with respect to the lowest energy ground state
for bilayer GaS and bilayer ZrS$_2$ as a function of the sliding parameter
(c) Band structure for a selected set of $\alpha$ values for bilayer GaS.
(d) Polarization evolution, in units of the spinless quantum of polarization, 
as a function of $\alpha$ for bilayer ZrS$_2$.}

\label{FIG-3}
\end{figure*}

For the second effect, the material we select is formed by two monolayers
of ZrS$_2$, which is an experimentally synthetized structure 
\cite{JACS_2015_ZrS2,APPsci_2016_ZrS2}. This material
has been found to exhibit excellent properties for electrocatalysis and 
optoelectronics
\cite{Chemmat_2019_ZrS2,2dmat_chemrev_2017,TMDs_Chemrev_2018,
ACS_PC_2024_ZrS2}.
This time, the structure is formed with the S atoms occupying 
a twofold WP as in the previous material GaS, and the Zr atoms from each
monolayer combine to conform another twofold WP. The lattice structures of
the three HSS are shown in FIG. \ref{FIG-3}a. Note that in this 
material the stackings are labelled by considering the
positions of the S atoms to maintain agreement with the
previous convention. Also, the order of the stackings in the 
cyclic process is A,B,C,A.

In this case, the simulation of the sliding
includes the ground state energy calculation along with
band structure calculations to check that the complete set of valence
bands remains gapped along the process. After that, a computation of the
polarization for the real space lattice directions $\mathbf{a}_1$ and
$\mathbf{a}_2$ is carried. The details for each step in this procedure are
given in the Appendix. The ground state calculations give the variation in
total energy with the sliding parameter, and we display the results of these
calculations in FIG. \ref{FIG-3}b.

In FIG. \ref{FIG-3}d the polarization interpolated from a set of
first-principles results for particular values of alpha (see Appendix
for details) along the direction $j=1$ as a function
of $\alpha$ is depicted for a complete adiabatic cycle. In analogous form to 
the minimal model, the polarization shows a winding in the $(\alpha,k_1,k_2)$
space. This confirms the adiabatic charge pump effect in this sliding bilayer.
The Chern number extracted from the FIG. \ref{FIG-3}d plot is $C(1)=-12$. 
For bilayer ZrS$_2$, the valence band manifold contains 24 bands. As we
previously discussed in section \ref{II.C}, the additive nature of the Chern 
number allows to predict its total magnitude by summing the contribution for
each minimal set of bands (a set of two bands in these systems). In this way,
as we have 24 bands then we obtain that $|C(j)|=24/2=12$, which agrees with the
first-principles result. The information of the sign of the Chern number
requires additional information on the number of bands. Since it depends on 
the value of the $j$ index and also on the direction of the sliding in real
space.  

\section{Conclusions}\label{IV}

In summary, we have studied the adiabatic sliding in bilayers composed
of one monolayer type with symmetries pertaining to the SG \# 164.
We focused on describing two effects that arise due to the sliding.
In the first place, a symmetry-enforced metal-insulator transition can be 
produced by connecting two stacks that exchange the WP, such that
the character of the bands change as we go from one stacking to the
other. A minimal 
four-band tight-binding model can readily describe the
transition, and we also show that a similar outcome
arises in a bilayer of the GaS material. 

Besides this effect, we found that a topological adiabatic charge
pump could be established when considering a cyclic sliding process in a
bilayer system that preserves its insulating character throughout
the operation. This pump can be characterized
by the evolution of the polarization along the real space lattice
vectors as a function of the sliding parameter. The topological
feature can be determined using the Chern number defined in 
$(\alpha,k_1,k_2)$ space. This Chern number is related to the
quantized transport properties that the system can show and scales
with the number of bands. 
The same minimal tight-binding model could spot this effect 
and was also shown to appear in the 
the bilayer form of ZrS$_2$, where a Chern number
of $C(j)=-12$ was encountered. 

Apart from the exemplary materials, the proposed effects may occur
in other materials. For instance, we have observed the metal-insulator
transition in bilayer AlS and
expect it to be present in bilayer GaO since these materials
have similar electronic band structure and filling as bilayer GaS
(see the database associated to ref. \cite{2D_CMDB_2021} for details).  
In the case of the charge pump, low spin-orbit materials with a
monolayer structure similar to ZrS$_2$ are good candidates. 
Potential candidates are the bilayer forms of TMDCs 
such as NiX$_2$, X=$\{$S,Se$\}$, and bilayer Mxenes such as
M$_2$O$_2$, M=$\{$Sc,Ti$\}$. 
As the model suggests, both phenomena are likely to be present
in many systems and, as such, could be realized experimentally 
in device form. Also, it would be interesting to explore 
potential extensions of the results here presented, 
such for example, how magnetism can affect the ground state
properties and transitions. 
Also, the impact of including elements with 
moderate to high spin-orbit coupling may be of interest.
Finally, the consequences of lowering the spatial
symmetry of the special stackings to structures with groups 
different from SG \#164 will likely produce additional
features associated with the symmetry-breaking process. 

\section*{Appendix: Computational details}\label{A}

The tight-binding model presented in section \ref{II} was 
numerically implemented using the Python package PythTB 
\cite{pythtb},
which can result in band structures and also has
implemented the Berry phase calculation as described in 
\cite{vanderbilt_2018}.
For the first-principles computations, we employed the Quantum 
ESPRESSO code (QE) \cite{QE_2009,QE_2017}. The calculations
contemplated first a structural relaxation for each sliding
stacking considered where the force tolerance was set to 
0.001 Ry/Bohr. These structures were employed to compute 
the ground state and the band structure for each material
and for each value of the parameter $\alpha$. An energy
cutoff of 80 Ry and a Monkhorst-Pack grid of
10$\times$10$\times$1 was used with the
Perdew-Burke-Ernzerhof (PBE) functional \cite{PBE_funct}
along with norm-conserving pseudopotentials 
\cite{Norm-cons_pseudo}
retrieved from the PseudoDojo library \cite{pseudoDojo}.
The energy tolerance was selected to be 10$^{-8}$ Ry. 

For the polarization calculation, the internal routine of 
QE was used, which follows the same approach as in the
tight-binding calculations \cite{polari_1993}. For this
particular step, 10 divisions for the k-line in reciprocal
space yield well-converged results. The calculation of this
polarization for the particular case of the ZrS$_2$ bilayer
was carried at values of $\alpha$ following the relation
$\alpha=n/24$ with $n$ an integer between 0 and 24. The rest of 
the points to construct the charge pump graph in FIG. \ref{FIG-3}d 
were obtained by interpolating between the calculated values
always keeping the modular character of polarization, that
restrict its values within 1 and -1, in units of the spin degenerate
quantum of polarization. 

\section*{Acknowledgements}
This work has been supported by the postdoctoral position from 
Universidad T\'ecnica Federico Santa Mar\'ia, 
and Chilean FONDECYT Grant 1220700.  

\newpage
\bibliography{ref}

\end{document}


\author{\text{Sergio Bravo}$^\dagger$}
\author{\text{P.A. Orellana}}
\author{\text{L. Rosales}}
\affil{Departamento de F\' isica, Universidad
T\'ecnica Federico Santa Mar\' ia, Valpara\' iso, Chile}

\affil[ ]{ }
\affil[$\dagger$]{sergio.bravo@usm.cl}

\date{}

\maketitle


\section*{Band representations for space group \#164}

\begin{table}[ht]
\centering
\begin{tabular}{@{}cccc@{}}
\toprule
WP & SSG irrep & orbitals & EBR at HSP   \\ \midrule
$2c$ & A$_{1}$ & $s$ , $p_{z}$, $d_{z^2}$  & $\Gamma_{1}^{+}$ - $K_1$ - $M_{1}^{+}$  \\
     &         &                           & $\Gamma_{2}^{-}$ - $K_2$ - $M_{2}^{-}$  \\
$2c$ & E       & ($p_{x}$,$p_{y}$),($d_{xy}$,$d_{yz}$),($d_{xy}$,$d_{x^2-y^2}$) 
& $\Gamma_{3}^{+}$ - $K_3$ - $M_{1}^{-} \oplus M_{2}^{+}$ \\
     &         &                                                                
     & $\Gamma_{3}^{-}$ - $K_3$ - $M_{1}^{-} \oplus M_{1}^{+}$ \\
$2d$ & A$_{1}$ & $s$ , $p_{z}$, $d_{z^2}$                                       
     & $\Gamma_{1}^{+}$ - $K_1$ - $M_{1}^{+}$                 \\
     &         &                                                                
     & $\Gamma_{2}^{-}$ - $K_2$ - $M_{2}^{-}$\\
$2d$ & E       & ($p_{x}$,$p_{y}$),($d_{xy}$,$d_{yz}$),($d_{xy}$,$d_{x^2-y^2}$) 
     & $\Gamma_{3}$ - $K_3$ - $M_{1}^{+} \oplus M_{2}^{+}$ \\
   &           &                                                                
   & $\Gamma_{3}^{-}$ - $K_3$ - $M_{1}^{-} \oplus M_{1}^{+}$ \\ 
   \bottomrule
\end{tabular}
\caption{elementary band representations (EBR) labelled for the space group \#164 high-symmetry points (HSP) 
in momentum space, induced from Wyckoff positions (WP) $2c$ and $2d$. The irreducible representation (irrep)
in real space for each WP and the corresponding site-symmetry group (SSG) is also featured.}
\label{tab:my-table}
\end{table}

\section*{Nearest-neighbors (NN) vectors for tight-binding model}

Intralayer first NN 
\begin{equation*}
    \delta_1 = (1/3,2/3,\Delta_{intra}); \hspace{1.5mm}
    \delta_2 = (1/3,-1/3,\Delta_{intra}); \hspace{1.5mm}
    \delta_3 = (-2/3,-1/3,\Delta_{intra}).
\end{equation*}
Here, $\Delta_{intra} = z_e-z_i$, where $z_e$ and $z_i$ are the vertical coordinates for external and internal atoms within one of the monolayers in the bilayer.\\
\\
Intralayer second NN
\begin{equation*}
    \gamma_1 = (1,0,0); \hspace{1.5mm}
    \gamma_2 = (0,1,0); \hspace{1.5mm}
    \gamma_3 = (1,1,0).
\end{equation*}
Interlayer internal NN
\begin{equation*}
    \xi_1 = (\alpha,-\alpha,2z_i); \hspace{1.5mm}
    \xi_2 = (\alpha,1-\alpha,2z_i); \hspace{1.5mm}
    \xi_3 = (-1+\alpha,-\alpha,2z_i).
\end{equation*}
Interlayer external NN
\begin{equation*}
    \zeta_1 = (2/3+\alpha,1/3-\alpha,2z_e); \hspace{1.5mm}
    \zeta_2 = (-1/3+\alpha,1/3+\alpha,2z_e); \hspace{1.5mm}
    \zeta_3 = (-1/3+\alpha,-2/3-\alpha,2z_e).
\end{equation*}

\newpage
\section*{Supplementary figures}

\begin{figure}[ht]
\includegraphics[width=\columnwidth,clip]{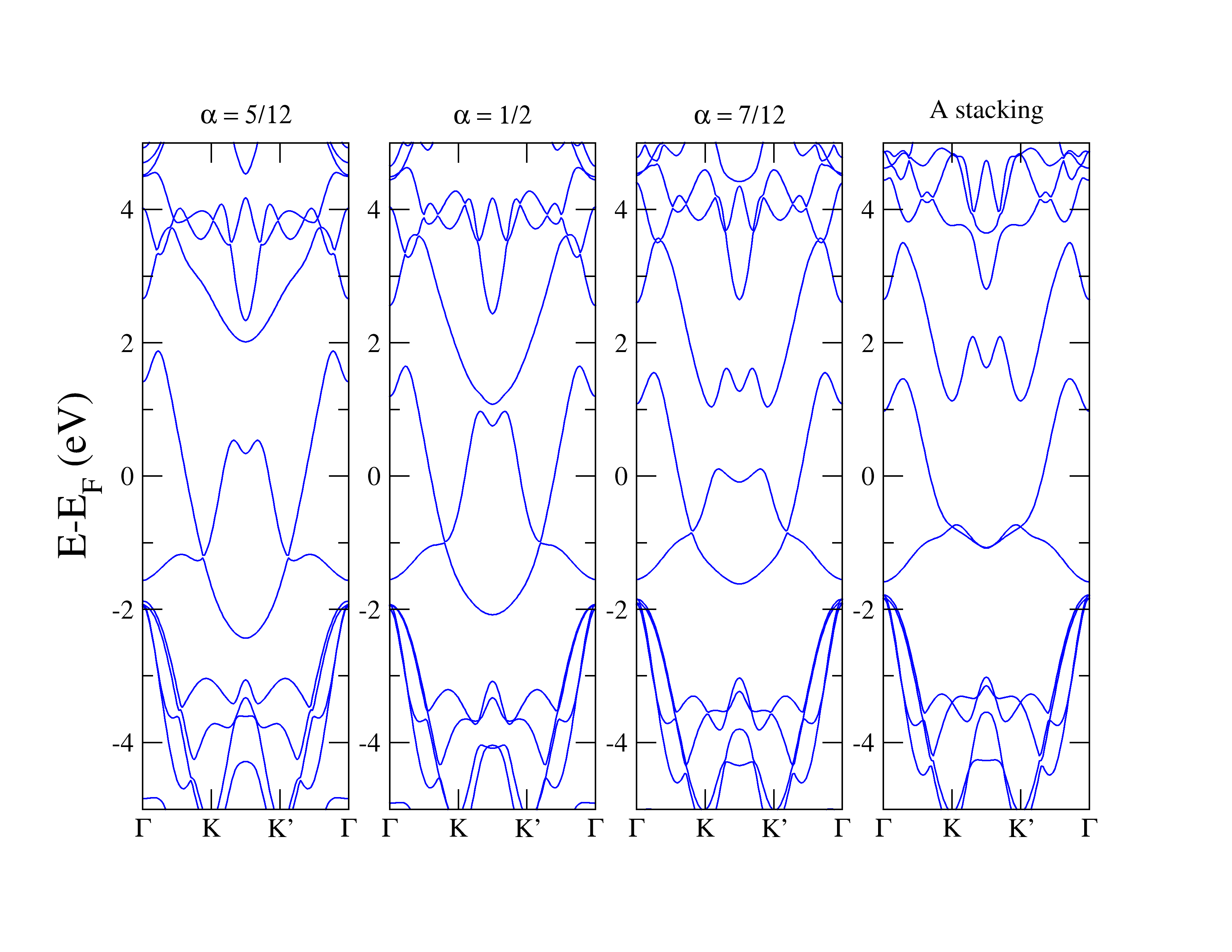}
\caption{Electronic band structure for values of the sliding parameter 
$\alpha$ within the range between staking C ($\alpha=1/3$) and 
stacking A ($\alpha=2/3$).}
\label{FIG-S1}
\end{figure}

\begin{figure}[ht]
\includegraphics[width=\columnwidth,clip]{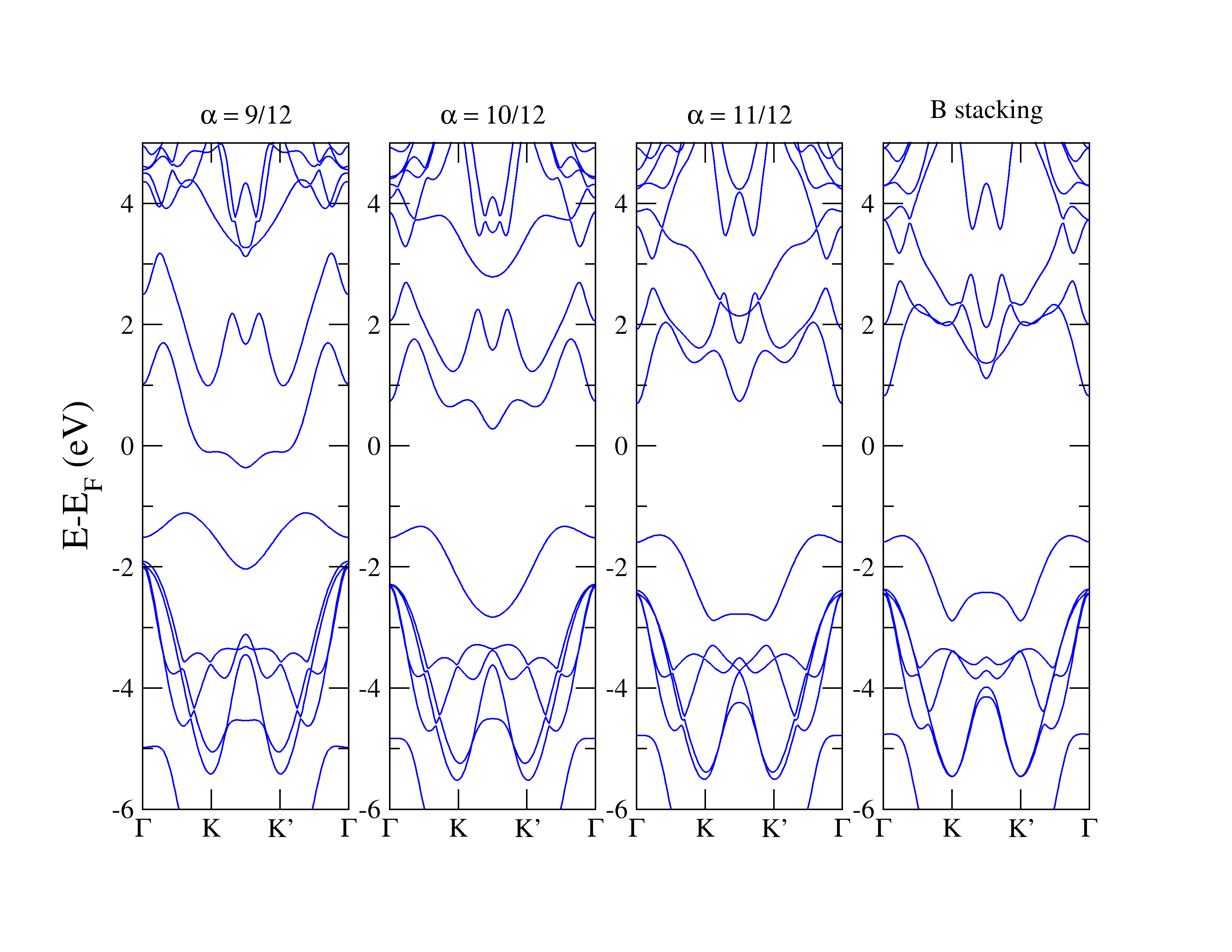}
\caption{Electronic band structure for values of the sliding parameter 
$\alpha$ within the range between staking A ($\alpha=2/3$) and 
stacking B ($\alpha=1$).}
\label{FIG-S2}
\end{figure}